# Geometry selective colossal negative dielectric permittivity in $CaFe_2O_4$ nanostructures

Sourav Sarkar*, and Kalyan Mandal*.

Department of Condensed Matter and Materials Physics, S. N. Bose National Centre for Basic Sciences, Sec III, Salt Lake, Kolkata 700106, India.

*To whom the correspondence should be given: sourav.sarkar@bose.res.in
kalyan@bose.res.in

**Abstract:**

   Negative permittivity metamaterial is a scientifically rich avenue due to its tremendous application in several arena of materials research including novel superlens, band-gap materials, invisibility cloaks, antenna and filter design. Traditionally, epsilon negative (ENG) behaviour is achieved in multi-phase composites with the addition of conducting metal fillers. However, this study reports colossal ENG feature in a single phase Calcium Ferrite for a particular nano hollow spherical (NHS) morphology, without the use of any filler. On the contrary, the same material synthesized in a different morphology, namely, nano solid sphere (NSS) shows conventional dielectric behaviour. Occurrence of ENG is successfully interpreted with the phase inversion of dominant polarization within the hollow cavity of NHS. This report marks a significant step in realizing colossal ENG in a single phase material just by restructuring the nanoscale morphology.

Electromagnetic metamaterials with negative permittivity and/or permeability have triggered tremendous curiosity among material scientists in recent decades. As their wave vector (k) follows left-handed rule, they are also called left-handed materials (LHM), theoretically proposed by Veselago in 1960s [1–3]. These LHMs exhibit several unconventional properties, such as, reversed Doppler effect [4], reversed Cherenkov radiation [5], negative refraction [6,7]. Due to these special traits, they are used in filter [8] and antenna design [9], invisibility cloak [10], perfect lens [11], electromagnetic shielding [12]. Ordered metamaterials are designed using artificially engineered periodic structures including double-split-ring and S-shaped resonators, metallic nanoclusters. However, their complex fabrication has driven growing interest in random metamaterials which can be synthesized through conventional chemical techniques.

The unique properties of metamaterials arise from epsilon-negative (ENG) ($\varepsilon<0$), Mu-negative (MNG) ($\mu<0$), or double negative (DNG) ($\varepsilon, \mu<0$) features [13], with ENG behaviour being particularly significant. Random metamaterials, or metacomposites, composed of multicomponent systems, offer ENG properties. Several studies have demonstrated this, including activated carbon felt-polyaniline [14], carbonized wood@Prussian blue derivative (CW@PBD) [12], Poly(butylene adipate-co-terephthalate)/FeCoNi@carbon systems [15], along with other composites such as Ag nanoparticle-immobilized carbon fiber [16], graphene/carbon black/$CaCu_3Ti_4O_{12}$ [17], carbon nanotube/$Bi@Fe_3O_4$ [18], polyvinyl alcohol–graphite [19], carbon nanofiber/poly(vinylidene fluoride) [20]. These systems typically consist of conductive fillers within a ceramic matrix and exhibit negative permittivity once the filler concentration exceeds the percolation threshold, below the plasma frequency of the metallic component. In contrast, fewer studies focus on achieving ENG behaviour in conventionally fabricated composites via nanoscale geometry [2]; for instance, ENG has been reported in cup-stacked carbon nanofibers embedded in polymer matrices.

However, achieving appreciable homogeneity in multicomponent systems remains challenging. A more effective approach is to realize ENG in single-phase materials, enabling better control and congruity [21,22]. Among several options, spinel ferrites ($MFe_2O_4$, M being metal cation) are promising candidates due to their easy synthesis, morphological versatility, tunable magnetic and dielectric properties, and room temperature stability [23–27]. Typically, ENG is reported for a particular morphology or composition of a sample. However, a systematic comparison of the same material across different morphologies, crucial for both fundamental and applied studies, remains largely unexplored.

Among spinel ferrites, alkaline earth (with M = Mg, Ca, Sr) systems are suitable candidates, supported by our earlier report of negative permittivity in $Mg_{0.3}Fe_{2.7}O_4$ [28], though the role of morphology towards ENG feature was not clearly addressed. Subsequent work on $MgFe_2O_4$ with a unique hollow spherical geometry highlighted morphological effects [29], but ENG appeared only over a narrow frequency range and much above room temperature. Therefore, it is important to realize ENG over a broad frequency spectrum near room temperature. This goal has been achieved in this work using Calcium Ferrite ($CaFe_2O_4$) with higher dielectric permittivity [30,31]. Accordingly, $CaFe_2O_4$ is synthesized in two morphologies: nano solid spheres (NSS) with compact spherical morphology, and nano hollow spheres (NHS) with a hollow central region surrounded by thick shell, and their dielectric properties are systematically compared. Among them, NHS shows colossal ENG which is explained from its unique geometry.

CaFe$_2$O$_4$ (CaFO) NSS and NHS are synthesized using a template free solvothermal method[23,24,29], as described in the Supplemental Material (SM). Impurity free single phase structure is confirmed for both NSS and NHS from the X-ray Diffraction patterns (Fig. S1, SM).

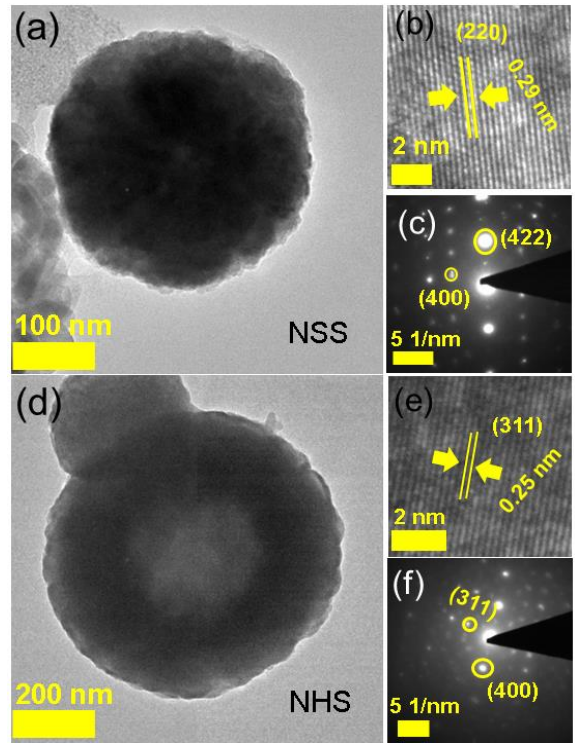

Fig. 1: TEM micrographs of CaFO (a) NSS, and (d) NHS. The thick shell and hollow cavity are visible for NHS. (220) and (311) crystallographic planes with the corresponding interplanar distances are shown for (b) NSS and (e) NHS, respectively. The SAED patterns identified with different crystallographic planes are shown for (c) NSS and (f) NHS.

Transmission Electron Micrographs (TEM) are presented in Fig. 1. NSS has solid spherical morphology, as confirmed in Fig. 1(a), and a broad size distribution with an average diameter of 163±65 nm (Fig. S3(a), SM). Alternatively, NHS shows a sharp contrast between the outer and inner regions, confirming a hollow cavity enclosed by thick shell (Fig. 1(d)). The average diameter of the cavity and shell thickness are 260±91 and 191±22 nm, respectively [Fig. S3(b, c), SM]. The separation between crystallographic planes observed in high-resolution TEM and Selected Area Electron Diffraction (SAED) patterns (for example, (220), (311), (400) and so on) match well with XRD analysis.

Figures 2(a) and (b) show the variation of AC conductivity ($\sigma_{ac}$) with the applied field frequency ($f$) for NSS and NHS, respectively, over the temperature ($T$) range 298 K-573 K (20 K intervals). For NSS [Fig. 2(a)], $\sigma_{ac}$ increases with $f$ up to around 2 MHz at all $T$, indicating typical insulating behaviour [18,32]. In this $f$-range, the $\sigma_{ac}$ vs. $f$ curves are fitted well with Jonscher's power law (black solid lines)[33,34]:

$$\sigma_{ac} = \sigma_{dc} + A\omega^n, \ldots(1)$$

where, $\sigma_{dc}$, $A$, $\omega$, and $n$ ($0<n<1$) respectively denote dc conductivity, pre-exponential factor, angular frequency of external electric field, and power-law exponent. This indicates hopping movement of charge carriers under the applied field. However, deviation from the law occurs above 2.5 MHz, as charge carriers can no longer follow the high frequency field. In contrast, Fig. 2(b) shows $f$ variation of $\sigma_{ac}$ for NHS which is 100 times higher than that of NSS. At lower $T$ up to 373 K, $\sigma_{ac}$ increases with $f$ according to equation (1) (cyan fitted curves). Above 373 K, $\sigma_{ac}$ remains constant up to certain $f$ and thereafter decreases due to much higher $f$, as discussed earlier.

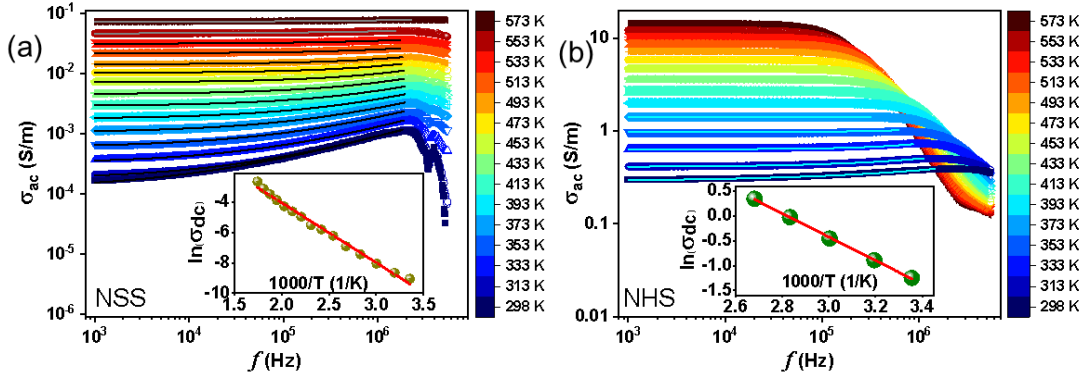

Fig. 2: Frequency ($f$) variation of ac conductivity of CaFO (a) NSS, and (b) NHS at different temperatures, T. The adjacent colour bar indicates the colour code for different T. The corresponding fits according to Jonscher's law (equation (1)) are shown in solid black or grey line for NSS and cyan line for NHS. The law is valid only up to certain $f$ for all $T$ for NSS and $T$ up to 373 K for NHS. Corresponding insets show Arrhenius plot (natural logarithm of dc conductivity vs. inverse of temperature). The solid red line indicates linear fitting.

For both the samples $\sigma_{ac}$ increases with $T$ due to enhanced thermal energy, allowing more frequent hopping and thus increased conductivity. The $\sigma_{dc}$ values obtained from fitting according to equation (1) for NSS and NHS at various $T$ are analysed through the Arrhenius model:

$$\sigma_{dc} = \sigma_0 \cdot e^{\frac{-E_a}{k_B T}} \quad \dots(2)$$

where, $\sigma_0$ is a constant, $E_a$ is the activation energy, and $k_B$ is the Boltzmann constant. This model captures the thermally activated overcoming of energy barriers, available to the charge carriers while movement through the sample. From the slope of the linear fit of $ln(\sigma_{dc})$ vs. $1000/T$ plots in Fig. 2(c) and (d), $E_a$ can be calculated. Faster hopping in NHS is evident from its significantly lower $E_a$ (0.2 eV) compared to that of NSS (0.34 eV).

The real part of relative permittivity ($\varepsilon'$) for NSS and NHS as a function of $f$ is depicted in Fig. 3(a) and (b), respectively, at various $T$. For NSS, $\varepsilon'$ decreases continuously with increasing $f$, consistent with Maxwell-Wagner theory of interfacial polarization. According to this theory, the material consists of conducting grains separated by insulating grain boundaries [29]. At high $T$ and low $f$, thermally activated charge carriers conduct through the grains and accumulate at grain boundaries, enhancing $\varepsilon'$. With the increase of $f$, $\varepsilon'$ reduces, approaching a negligible constant value. This behaviour is suppressed at lower $T$ due to reduction in thermal energy and hence the movement of charges. The inset of Fig. 3(a) shows the enlarged portion of the curve near $\varepsilon' = 0$, confirming positive $\varepsilon'$ throughout the whole $f$ and $T$.

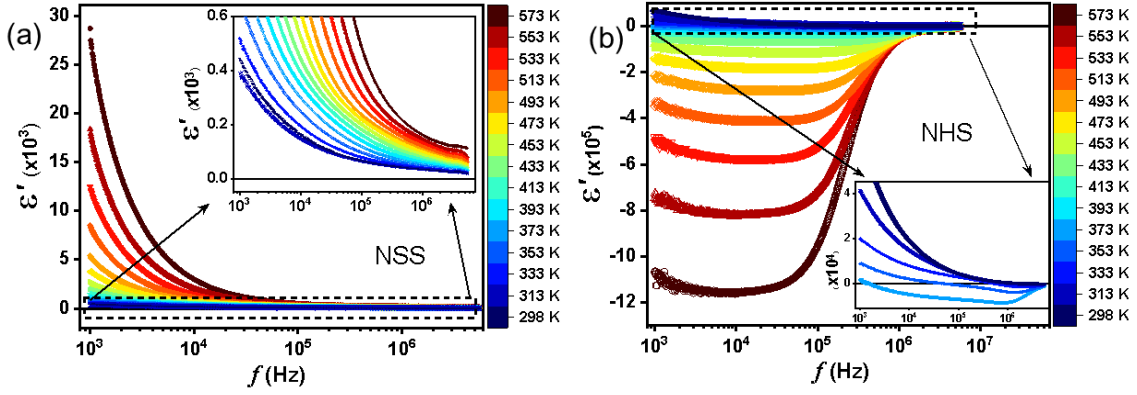

Fig. 3: Frequency variation of real part of relative permittivity, $\varepsilon'$, of CaFO (a) NSS, and (b) NHS at different temperatures, $T$. The adjacent colour bar indicates the colour code for different $T$. The inset of (a), and (b) show the enlarged portions of the corresponding curves for NSS and NHS, respectively, near $\varepsilon' = 0$. A solid black horizontal line at $\varepsilon' = 0$ is drawn for a guide to the eye.

In contrast, NHS exhibits anomalous $f$-dependence of $\varepsilon'$ at different $T$ [Fig. 3(b)]. The inset of Fig. 3(b) shows the crossover of $\varepsilon'$ from positive to negative value within the $T$ range of 298 K to 373 K. Above 373 K, NHS shows colossal negative $\varepsilon'$ (the magnitude being ~$10^4$-$10^5$) throughout the entire $f$-range and approaches to 0 at higher $f$. Very high negative value of $\varepsilon'$ can be interpreted from its hollow-spherical geometry. Therefore, different polarization mechanisms in the two samples are schematically shown in Fig. 4. Under the application of an electric field, $\vec{E}$, a polarization, $\vec{P}$, is developed in a material. For NSS, the direction of $\vec{P}$ is along the direction of $\vec{E}$, as shown in Fig. 4(a). On the other hand, in presence of $\vec{E}$, induced charges in NHS develop polarization in the shell region ($\vec{P_s}$) along the direction of $\vec{E}$, whereas the direction of induced polarization inside the hollow cavity, $\vec{P_c}$, is opposite to $\vec{E}$, as shown in Fig. 4(b). As the central cavity has a smaller radius than that of the outer surface of NHS, the surface charge density corresponding to the inner cavity will be larger than that of the outer shell region for a fixed amount of induced charges. Additionally, the cavity contains only air with extremely low electrical conductivity ($10^{-12}$ to $10^{-15}$ S/m)[35–37] in this $f$-region, while NHS has conductivity of 1-10 S/m, as shown in Fig. 2(b). Therefore, almost $10^{12}$ times higher conductivity and leakage reduce the charge storage and polarization in the shell area. In contrast, higher amount of charges get stored at the cavity boundary. Hence, a much stronger $\vec{P_c}$ for the inner cavity will be achieved than that of the shell ($\vec{P_s}$). Therefore, a net $\vec{P}$, opposite to $\vec{E}$ will develop. With the increase in $T$, accumulation of charge will be enhanced due to higher mobility coming from the gain in thermal energy. As a result, phase inversion and colossal negative $\varepsilon'$ (Fig. 3(b)) is achieved in NHS. With much higher $f$, charge carriers can no longer follow $\vec{E}$, driving $\varepsilon'$ to approach 0. As the resultant $\vec{P}$ in NHS is opposite to applied field $\vec{E}$, the internal field developed in the hollow cavity adds to $\vec{E}$. Consequently, enhanced hopping increases the conductivity. Increasing $T$ enhances the internal field even more, and hence the conductivity [Fig. 2(b)]. Comparison with various composites (Table S1, SM) [38–46] shows that CaFO NHS exhibits one of the highest negative $\varepsilon'$ values reported for a single-phase material. While Drude like ENG behaviour has been previously observed in multi-composite systems with conducting fillers at hundreds of MHz [18,32,47–50], ENG in this study is achieved in highly dielectric CaFO through nanoscale morphology engineering without the use of any fillers in the low $f$ range (kHz-MHz).

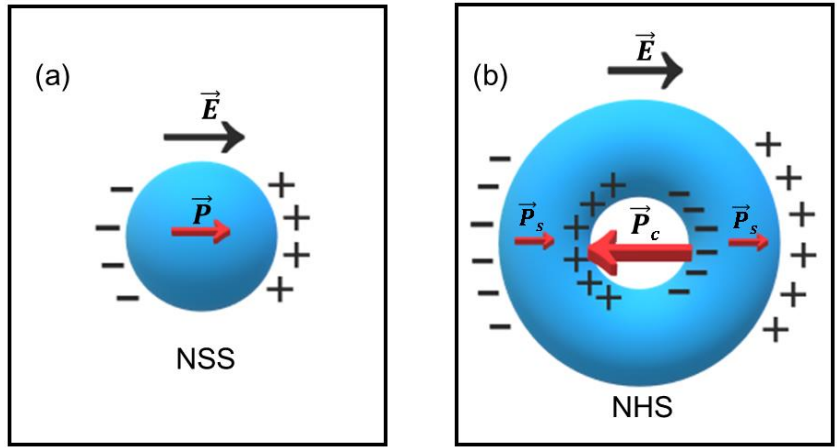

Fig. 4: Schematic interpretation of the corresponding polarization mechanism for (a) NSS and (b) NHS. The black and red arrows indicate the direction of the applied electric field and the polarization vector developed in the material. The larger arrow for polarization in central cavity ($\vec{P_c}$) than that of the shell region ($\vec{P_s}$) indicates stronger polarization in the cavity than the shell area for NHS.

To get deeper insights about the ENG behaviour in NHS, $f$ variation of the phase shift angle, $\Phi$ is analyzed. For clarity, $\Phi$ corresponding to some select temperatures only are depicted in Fig. 5 to show the overall trend. For NSS, $\Phi$ shows negative values throughout the whole $f$-range upto the highest $T$ [Fig. 5(a)]. This is an indication of the voltage lagging the current through the sample. The $\Phi$ at 298 K approaches -90º, signifying a purely capacitive behaviour. On the other hand, $\Phi$ for NHS indicates opposite behaviour [Fig. 5(b)]. For 298 K and 373 K, $\Phi$ shows negative values up to 1.5 MHz and 1.5 kHz, respectively, and then changes to positive value. For higher $T$, it shows positive values throughout the whole $f$-range. The positive value of $\Phi$ in NHS signifies development of polarization in the central cavity with opposite polarity, as shown in Fig. 4(b). This provides a phase interpretation of the ENG behaviour in NHS. With increasing $T$, $\Phi$ increases. Hence, the increasing ENG behaviour in NHS with $T$ can also be justified from the $\Phi$ variation.

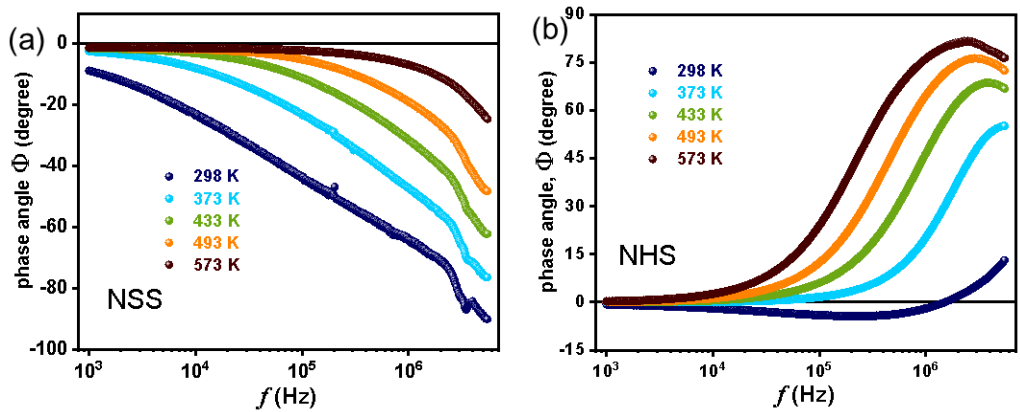

Fig. 5: Frequency variation of phase shift angle, $\Phi$, of CaFO (a) NSS, and (b) NHS at different temperatures, $T$. For clarity, $\Phi$ corresponding to some select $T$ only are shown, as indicated by the colour code. A solid black horizontal line at $\Phi = 0$ is drawn for a guide to the eye.

This study explores how a conventional single phase spinel ferrite can be fabricated in a particular nanostructure to achieve negative permittivity. CaFO in a unique nano hollow spherical morphology (NHS) shows epsilon negative (ENG) behaviour, whereas the same material in nano solid spherical morphology (NSS) shows conventional dielectric behaviour. The realization of ENG in single phase CaFO NHS without any filler finds its origin purely in its unique morphology. The phase inversion of polarization within the central cavity of NHS has been explained schematically. Additionally, measured value of phase shift angle supports this notion. Higher ac conductivity in NHS compared to NSS is further explained with this approach. The interesting findings of this work are expected to open new possibilities in realizing metamaterial traits by morphology engineering at the nanoscale.

**Acknowledgement:** Sourav Sarkar gratefully acknowledge INSPIRE, Department of Science and Technology (IF200252), Government of India for financial assistance. The authors would like to acknowledge the Technical Research Centre of SNBNCBS. The authors sincerely acknowledge the scientific discussions with Mr. Sk Kalimuddin, IACS, Mr. Deep Singha Roy, IACS, Dr. Priyanka Saha, Chalmers University of Technology, and Dr. Sudipta Chatterjee, Princeton University.

**Conflict of Interest:** The authors declare no competing interest.

**Date Availability statement:** All the data supporting the findings of this study are available within the article and its Supporting Information. Raw data that support the findings of the study are available from the corresponding author(s), upon reasonable request.

**Supplemental Material:** Detailed experimental procedures, X-ray diffraction patterns, FESEM images, Sze distribution curves, comparison table (Table S1), $\varepsilon''$ vs. $f$ curves along with other relevant plots are shown.